\documentclass{jfm}
\pdfoutput=1

\usepackage[pdftex]{graphicx}

\usepackage{natbib}
\usepackage{epstopdf}
\usepackage{bm}
\usepackage{amssymb}
\usepackage{amsbsy}
\usepackage{color}
\usepackage{mathrsfs}
\usepackage{amsmath}
\usepackage{amsfonts}
\usepackage{amssymb}
\usepackage{amsbsy}

\usepackage{color}
\definecolor{darkgreen}{rgb}{0,0.5,0}

\usepackage{mathrsfs}
\usepackage{graphicx}

\title[Drops on soft solids: Free energy and double transition of contact angles]{Drops on soft solids: Free energy and double transition of contact angles}

\author[L.A. Lubbers, J.H. Weijs, L. Botto, S. Das, B. Andreotti and J.H. Snoeijer]%
{L.\ls A.\ns L\ls U\ls B\ls B\ls E\ls R\ls S$^1$,\ns
J.\ls H.\ns W\ls E\ls I\ls  J\ls S$^1$,\ns 
L.\ns B\ls O\ls T\ls T\ls  O$^2$,\ns \break
S.\ns D\ls A\ls S$^3$,\ns 
B.\ns A\ls N\ls D\ls R\ls E\ls O\ls T\ls T\ls I$^4$,\ns 
\and J.\ls H.\ns S\ls N\ls O\ls E\ls I\ls J\ls E\ls R$^{1,5}$}

\affiliation{$^1$ Physics of Fluids Group, Faculty of Science and Technology, Mesa+ Institute, University of Twente,
7500~AE Enschede, The Netherlands.\\
$^2$ School of Engineering and Materials Science, Queen Mary University of London, E1 4NS, London, UK \\
$^3$ Department of Mechanical Engineering, University of Maryland, College Park, Maryland 20742, United States \\
$^4$ Physique et M\'ecanique des Milieux H\'et\'erog\`enes, UMR
7636 ESPCI -CNRS, Univ. Paris-Diderot, 10 rue Vauquelin, 75005, Paris, France\\
$^5$ Department of Applied Physics, Eindhoven University of Technology, P.O. Box 513, 5600~MB Eindhoven, The Netherlands.
}

\pubyear{??}
\volume{??}
\pagerange{??--??}
\date{?? and in revised form ??}
\begin{document}

\maketitle

\begin{abstract}
The equilibrium shape of liquid drops on elastic substrates is determined by minimising elastic and capillary free energies, focusing on thick incompressible substrates. The problem is governed by three length scales: the size of the drop $R$, the molecular size $a$, and the ratio of surface tension to elastic modulus $\gamma/E$. We show that the contact angles undergo two transitions upon changing the substrates from rigid to soft. The microscopic wetting angles deviate from Young's law when  $\gamma/Ea \gg 1$, while the apparent macroscopic angle only changes in the very soft limit $\gamma/ER \gg 1$. The elastic deformations are worked out in the simplifying case where the solid surface energy is assumed constant. The total free energy turns out lower on softer substrates, consistent with recent experiments.
\end{abstract}

\section{Introduction}
A liquid drop can deform a soft elastic substrate due to capillary forces \citep{L61, Rusanov75, R78,Shanahan87,bookDeGennes,PBBB08}. 
Elastic deformations take place over a length on the order of the elastocapillary length $\gamma/E$, where $\gamma$ is the liquid surface tension and $E$ the solid Young's modulus. 
Recent experiments have considered liquids on very soft elastomers with $\gamma/E$ of the order of 1-100 microns and have reported many interesting features, such as the geometry near the contact line \citep{PericetCamara08, JXWD11,Style13}, compression of the solid \citep{MDSA12}, evaporation and spreading dynamics \citep{CGS96,LGBGDBB07,PericetCamara08,SARLBB10}, as well as migration of droplets on substrates with a stiffness gradient \citep{StylePNAS13}.

While the contact angle of a drop on a  rigid, homogeneous substrate is governed by Young's law, the contact angle selection for a drop on a soft deformable substrate is still debated. One of the earliest theoretical approaches consisted of treating the elastic energy stored below the contact line as an effective line tension \citep{Shanahan87,White03,Style12}. This predicts an increase of the apparent, macroscopic contact angle on soft surfaces, which has been contradicted by recent experiments \citep{Style13}. The line tension approach did not include the surface energy of the solid, which turns out to be a crucial factor for the shaping of soft solids.  Such surface effects were introduced by \citet{JXWD11} and \citet{Limat12} in a purely macroscopic theory based on the balance of forces exerted ``on" the contact line. 
In this framework, the microscopic contact angles always obey Neumann's law, regardless of $E$, as if the substrate at the contact line was a liquid. 
The same work reveals that there exists a transition for the apparent macroscopic contact angle, controlled by the dimensionless parameter $\gamma/ER$, where $R$ is the drop size. 
By contrast, a microscopic description based on van der Waals interactions \citep{MDSA12b} suggests that Young's law for the microscopic contact angle is recovered for $\gamma/Ea \ll 1$, where $a$ is the characteristic length of molecular interactions. The formation of a solid cusp below the contact line arises for $\gamma/Ea \gg 1$, and the corresponding contact angles generically differ from Neumann's law, even for very soft substrates.

The controversy on the selection of contact angles on soft substrates underlines the difficulty of defining a force balance near the contact line \citep{MWSA11}. This has two distinct reasons:  On the one hand, the difference between macroscopic and microscopic descriptions of capillarity, and on the other hand, the difference between surface stress $\Upsilon$ (force per unit length) which is manifested in the superficial layers of a solid, and surface free energy $\gamma$ (energy per unit area). The latter is due to the coupling of elastic strain and surface free energy, an effect that is absent for liquid-liquid interfaces. Surface stress and surface energy are related by a thermodynamic law known as the Shuttleworth equation, $\Upsilon = \gamma + \frac{d\gamma}{d\epsilon_\parallel}$, where $\epsilon_\parallel$ is the elastic strain parallel to the interface \citep{Shuttleworth50}. The strain dependence $d\gamma/d\epsilon_\parallel$ is directly responsible for tangential elastic stress transmitted below the contact line \citep{DMAS11,MDSA12,WAS13}, usually ignored in elasto-capillary modeling. 

In this paper we revisit the contact angle selection from a thermodynamic perspective, using variational calculus -- i.e. without relying on a mechanical view in terms of a force balance. The challenge is to derive the contact angles directly by minimising capillary and elastic free energies. An important question then is whether microscopic and macroscopic descriptions of capillarity will give consistent results. To make progress, we restrict ourselves to the case where the Shuttleworth-effect is absent, i.e. $d\gamma/d\epsilon_\parallel=0$, for which it was previously shown that the tangential elastic stress vanishes \citep{WAS13}. Combined with an incompressible thick substrate (Poisson ratio $\nu=1/2$), this implies no in-plane displacements, making the problem amenable for detailed variational analysis.

\begin{figure}
\includegraphics{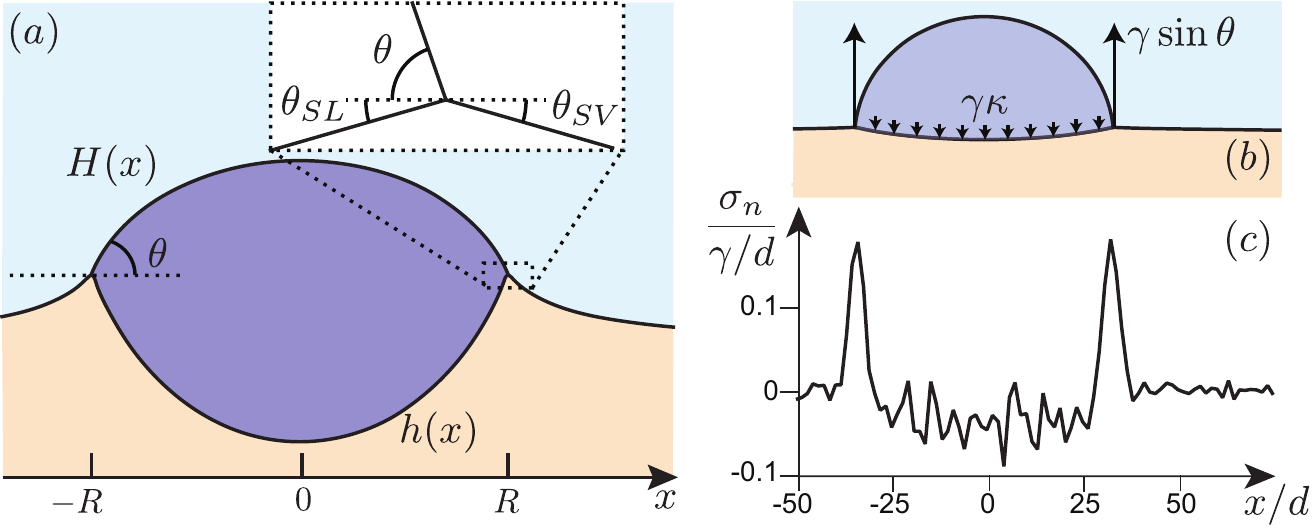}\label{fig:sketch}
\caption{(a) The liquid and solid surface shapes are described by $H(x)$ and $h(x)$ respectively. The contact line geometry is characterized by the angles $\theta$, $\theta_{SV}$ and $\theta_{SL}$ (see inset). (b,c) Normal component of capillary stress exerted on the solid. (b) Macroscopic view given by  Eq.~(\ref{eq:fn}). A force per unit length is pulling on the solid at the contact line positions while a Laplace pressure is applied below the drop.  (c) Microscopic view of the normal stress $\sigma_n$, below a Lennard-Jones nanodroplet  in Molecular Dynamics (adapted from \citet{WAS13}). The force near the contact lines in spread over a finite width $a$, of a few molecular sizes (units in terms of molecular size $d$). At the center of the drop one recognizes the slightly negative capillary stress.}
\end{figure}

The key questions resolved here are how the contact angles and the elasto-capillary free energies evolve upon changing the substrates from perfectly rigid (no deformation) to extremely soft (no elasticity). We predict the angles $\theta$, $\theta_{SV}$ and $\theta_{SL}$ as defined in Fig.~\ref{fig:sketch}a, as a function of the ``softness parameters" $\gamma/Ea$ and $\gamma/ER$. We show that both parameters govern a transition from Young's law to Neumann's law, but apply to different features of the angles in Fig.~\ref{fig:sketch}. This reveals the connection between previously proposed results in a single framework. Finally, the full shapes of the drops and the surface deflections are worked out for a case where the solid surface energies $\gamma_{SV}=\gamma_{SL}$. We find that the free energy is lower on softer substrates, and relate this to experiments of drop motion on substrate exhibiting a stiffness gradient \citep{StylePNAS13}.
 

%

\section{Equilibrium conditions from variational analysis}
\subsection{Macroscopic theory of capillarity}
\subsubsection{A single contact line}
Considering a semi-infinite inompressible solid ($\nu=1/2$) in a two-dimensional system, the elastic free energy per unit length can be expressed as a surface integral:
\begin{equation}
\label{eq:Fel}
\mathcal{F}_{el} =  \frac{E}{2} \int_{-\infty}^\infty \frac{dq}{2\pi} \, \widehat{Q}(q) \left[ \hat{h}(q)\hat{h}(-q) +\hat{u}(q)\hat{u}(-q)  \right],
\end{equation}
where $\hat{h}(q)$ and $\hat{u}(q)$ are Fourier transforms of the normal and tangential displacements, $h(x)$ and $u(x)$, at the interface  \citep{LAL96}. The Green's function reads $\widehat{Q}(q)= 2|q|/3$ in the incompressible limit.

Before analyzing a two-dimensional drop (Fig.~\ref{fig:sketch}), we consider the simpler case of a single contact line located at $x=R$. The liquid-vapor interface is described by $H(x)$, while $h(x)$ is the profile of the solid surface. When useful, we will use  indexed notation $h_{SL}$ or $h_{SV}$ to distinguish if the solid is wet or dry, with surface free energies $\gamma_{SL}$ and $\gamma_{SV}$. The total surface free energy in this case is
\begin{equation}
\label{eq:Fc}
\mathcal{F}_c=  \gamma \int_{-\infty}^R dx  \left(1+H'^2\right)^{1/2} + \gamma_{SL}\int_{-\infty}^R dx  \left(1+h_{SL}'^2\right)^{1/2} 
+ \gamma_{SV} \int_R^{\infty} dx  \left(1+h_{SV}'^2\right)^{1/2},
\end{equation}
 where $\infty$ refers to a very large (finite) distance far away from the contact line. We anticipate a slope discontinuity $h'_{SV}\neq h'_{SL}$ at $x=R$. The condition that $H$, $h_{SL}$ and $h_{SV}$ take the same value at the contact line can be enforced by introducing two Lagrange multipliers, $\lambda_{SL}$ and $\lambda_{SV}$, into the energy functional:
\begin{equation}
\mathcal{F}_{tot} = \mathcal{F}_{el} +\mathcal{F}_c + \lambda_{SL} \left[ H(R) - h_{SL}(R) \right] +\lambda_{SV} \left[ H(R) - h_{SV}(R) \right],
\end{equation}
where the Lagrange multipliers can be interpreted as reaction forces (per unit length). Minimization must be performed with respect to $H(x)$, $h(x)$, and the contact line position $R$. As explained in the introduction, the  tangential displacements $u(x)$ vanish when the surface free energies are assumed constants, i.e. independent of the elastic strain.

First, we consider the variation $\delta H(x)$, which gives

\begin{eqnarray}\label{eq:dH}
\delta \mathcal{F}_{tot} = \delta H(R) \left[\lambda_{SL} + \lambda_{SV} +  \frac{\gamma H'}{\left( 1+H'^2\right)^{1/2}} \right]  
+ \int_{-\infty}^R dx \, \delta H(x) \left[- \gamma \kappa \right].
\end{eqnarray}
The boundary term $\delta H(R)$ originates from the constraints on the contact line position and from the integration 
by parts of the capillary term $\mathcal{F}_c$ with respect to $\delta H'$ and gives $\lambda_{SL}+\lambda_{SV} = \gamma \sin \theta$.
The second term on the right-hand side involves the curvature $\kappa = H''/(1+H'^2)^{3/2}$, which  is zero for a single contact line. 

Second, we consider the variation $\delta h(x)$, which gives
\begin{eqnarray}\label{eq:varh}
\delta \mathcal{F}_{tot} &=& \delta h_{SL}(R) \left[-\lambda_{SL} + \frac {\gamma_{SL}h_{SL}'}{\left( 1+h_{SL}'^2\right)^{1/2}} \right]  
+ \delta h_{SV}(R) \left[-\lambda_{SV} - \frac{ \gamma_{SV}h_{SV}'}{\left( 1+h_{SV}'^2\right)^{1/2}} \right]  \nonumber \\
&&
+ \int_{-\infty}^R dx \,\delta h_{SL}(x) \left[- \gamma_{SL} \kappa_{SL} \right] 
+ \int_{R}^\infty dx \,\delta h_{SV}(x) \left[- \gamma_{SV} \kappa_{SV} \right] \nonumber \\
&&
+ \int_{-\infty}^\infty dx \, \delta h(x) \left[ E \int_{-\infty}^\infty \frac{dq}{2\pi}  \widehat{Q}_n(q) \hat{h}(q) e^{iqx} \right],
\end{eqnarray}
where the first four terms in brackets are analogous to those in (\ref{eq:dH}), but they now express vertical forces acting on a corner of solid. The last term represents the inverse Fourier transform of the variation of $\mathcal{F}_{el}$, and involves the elastic normal stress
\begin{equation}\label{eq:defsigma}
\sigma_n(x) \equiv E \int_{-\infty}^\infty \frac{dq}{2\pi}  \widehat{Q}_n(q) \hat{h}(q) e^{iqx}.
\end{equation}
%
When $\sigma_n$ contains a Dirac $\delta$-function at $x=R$, this elastic stress could contribute to the boundary condition at $x=R$, but
within the macroscopic framework this can be excluded on a mathematical ground. As  $\widehat{Q} \sim |q|$, a $\delta$-function contribution  would require a (weakly) singular displacement, $h \sim \log|x-R|$, and hence a (weakly) diverging elastic free energy. 
The equilibrium condition at the contact line is therefore determined by the surface energies only \citep{JXWD11,Limat12}, implying a weaker singularity in the form of a discontinuity in $h'$. One can show that this implies a weakly diverging stress, $\sigma_n \sim \log|x-R|$, but an integrable free energy. Given this singularity, the physics may be significantly different when including microscopic effects. 

Following for now the macroscopic derivation, the variations at the contact line, $\delta h(R)$, determine the Lagrange multipliers,  $\lambda_{SL} = \gamma_{SL} \sin \theta_{SL}$, and $\lambda_{SV} = \gamma_{SV} \sin \theta_{SV}$. Combined with the boundary condition resulting from (\ref{eq:dH}), these expressions give 
\begin{equation}\label{eq:verneumann}
\gamma \sin \theta =  \gamma_{SL} \sin\theta_{SL} + \gamma_{SV} \sin\theta_{SV},
\end{equation}
which can be recognized as the vertical component of the Neumann condition. This condition is analogous to that describing a liquid lens floating on another liquid, since the elastic energy does not give a contribution at the contact line.
The mechanical equilibrium away from the contact line can be obtained by combining the integrals in (\ref{eq:varh}), giving
\begin{eqnarray}\label{eq:sigman}
\sigma_n(x)  &=& \gamma_{SL} \kappa_{SL} \Theta(R-x) + \gamma_{SV} \kappa_{SV}\Theta(x-R) 
\equiv  \gamma_s(x) \, \frac{h''}{\left( 1+h'^2\right)^{3/2}},
\end{eqnarray}
where $\Theta(x)$ is the Heaviside step-function, and $\gamma_{s}(x)$ the solid surface tension on the respective domains. This expression is not defined at $x=R$, where instead it is replaced by a boundary condition (\ref{eq:verneumann}). 
For $x\neq R$, the stress $\sigma_n$ balances the Laplace pressure due to curvature $\kappa_{SL}$ ($\kappa_{SV}$) of the wet (dry) solid interface.

We now consider variations $\delta R$ of the contact line position. This variation receives contributions from the integration limits in $\mathcal{F}_c$, from the contact line constraints (terms involving $\lambda_{SL}+\lambda_{SV}$), but once again not from the energy contribution $\mathcal{F}_{el}$. The total variation (to be evaluated at $x=R$) gives
\begin{eqnarray}
\delta \mathcal{F}_{tot} &=& \delta R \left\{  \gamma \left(1+H'^2\right)^{1/2} + \gamma_{SL} \left(1+h_{SL}'^2\right)^{1/2} -\gamma_{SV} \left(1+h_{SV}'^2\right)^{1/2}\right. \nonumber \\
&&\left. + (\lambda_{SV}+\lambda_{SL}) H'  - \lambda_{SL} h_{SL}'  - \lambda_{SV} h_{SV}' 
\right\} \nonumber \\
&=& \delta R \left\{ \frac{\gamma}{\left(1+H'^2\right)^{1/2}} + \frac{\gamma_{SL}}{\left(1+h_{SL}'^2\right)^{1/2}} 
-\frac{\gamma_{SV}}{\left(1+h_{SV}'^2\right)^{1/2}} \right\},
\end{eqnarray}
where we inserted the values for  $\lambda_{SL}$, $\lambda_{SV}$ obtained previously. In the terms of the angles indicated in Fig.~\ref{fig:sketch}, this expression gives the horizontal Neumann condition:
\begin{equation}\label{eq:horneumann}
\gamma \cos \theta + \gamma_{SL}\cos \theta_{SL} = \gamma_{SV} \cos \theta_{SV}.
\end{equation}
To summarize, we recover the usual Laplace pressure condition for the liquid interface $H(x)$. The solid interface shape $h(x)$ follows from the balance (\ref{eq:sigman}) of elastic stress $\sigma_n$ and solid Laplace pressure. The problem is closed by the vertical and horizontal Neumann conditions (\ref{eq:verneumann},\ref{eq:horneumann}), serving as boundary conditions for the angles at the contact line. 

\subsubsection{A two-dimensional drop}
We now consider a two-dimensional drop with contact lines at $x=R$ and $x=-R$.
The expression for the free energy is similar to that for a single contact line, except that we have to consider two dry domains, $x > R$ and $x< -R $, and one wet domain $-R < x < R$. 
In addition, minimization is done at constant drop volume $V$. This constraint is enforced by adding the Lagrange multiplier term 
\begin{equation}
\mathcal{F}_V = P \left[V - \int_{-R}^R dx\, \left( H(x) - h(x)\right) \right] 
\end{equation}
to the total energy. Since $H(x)=h(x)$ at $x=\pm R$, the volume constraint does not contribute to the boundary conditions, and we recover (\ref{eq:verneumann},\ref{eq:horneumann}). 
 It does, however, introduce a contribution in the equilibrium equations for $H(x)$ and $h(x)$, namely:

\begin{eqnarray}
-\gamma \kappa = P, \quad 
\sigma_n(x)  = - P \Theta(R-|x|) + \gamma_s(x) \, \frac{h''}{\left( 1+h'^2\right)^{3/2}}.
\end{eqnarray}
The first relation requires the liquid interface to have a constant curvature, corresponding to the Laplace pressure $P=\gamma \sin \theta/R$. The second equation shows that $P$ also acts as external stress on the boundary of the elastic medium below the drop. It is instructive to incorporate the vertical boundary condition (\ref{eq:verneumann}) in the elastic stress, as
\begin{eqnarray}\label{eq:sigmadrop}
\sigma_n =   \gamma \sin \theta f_n(x) + \left(\frac{ \gamma_s(x) h'}{\left( 1+h'^2\right)^{1/2}} \right)',
\end{eqnarray} 
where $f_n(x)$ is the distribution of normal stress due to the capillary forces,
\begin{equation}\label{eq:fn}
f_n(x) =  \delta(x-R) + \delta(x+R) - \frac{1}{R} \Theta(R-|x|).
\end{equation}
This corresponds to the classical mechanical view of two localized forces pulling upwards at the contact line, and a Laplace pressure pushing downward below the drop (Fig.~\ref{fig:sketch}b).

\subsection{Microscopic theory of capillarity}

In the limit of perfectly rigid solids one would expect to recover Young's law. However, the macroscopic theory derived above shows that the wetting angles are given by the Neumann conditions \emph{regardless of stiffness}, which is clearly unphysical. Also, the elastic stress was found to be singular at the contact line. These artefacts  are due to the assumption of localized line forces represented by $\delta$-functions in (\ref{eq:fn}). Because the interface is diffuse at the molecular level and due to the finite range of van der Waals interactions, capillary forces are in reality spread out over a finite width $a$, typically a few nanometers. This is illustrated in Fig.~\ref{fig:sketch}c, showing the normal stress exerted by a nanodroplet on an elastic substrate as measured in a Molecular Dynamics simulation \citep{WAS13}. 

The finite width of the interface can be taken into account in a truly microscopic description, using a disjoining pressure \citep{White03} or Density Functional Theory \citep{DMAS11,MDSA12b}. Here we propose a simplified model that captures the essence of these microscopic models, but yet allows  for a numerical solution with realistic separation of microscopic scale ($a\sim$~nm) and macroscopic scale ($R\sim$~mm).  The approach consists in retaining the boundary condition (\ref{eq:horneumann}), but representing the line forces in (\ref{eq:fn}) by a function $g$ of finite width
\begin{equation}\label{eq:fnmicro}
f_n(x) = \frac{1}{a} g\left(\frac{x-R}{a}\right) +\frac{1}{a} g\left(\frac{x+R}{a}\right) - \frac{1}{R} \Theta(R-|x|) .
\end{equation}
In our numerical solutions, $g$ is chosen as a normalized Gaussian, but other choices give similar results. With this approach, the contact angles display two distinct transitions that are governed by the parameters $\gamma/(Ea)$ and $\gamma/(ER)$.

\subsection{Dimensionless equations and solution strategy}

Solving for $h(x)$ from (\ref{eq:sigmadrop}) is challenging, owing to the integral nature of $\sigma_n$, the discontinuity in $\gamma_s(x)$, and the presence of non-linear terms. 
We simplify the problem by considering the case $\gamma_{SV}=\gamma_{SL}=\gamma_s$, and assume $h'^2 \ll 1$ to linearize the last term in (\ref{eq:sigmadrop}); the latter condition is enforced by choosing moderate values of $\gamma/\gamma_s$. We then solve the resulting equation by Fourier transform. 
\begin{figure}
\includegraphics{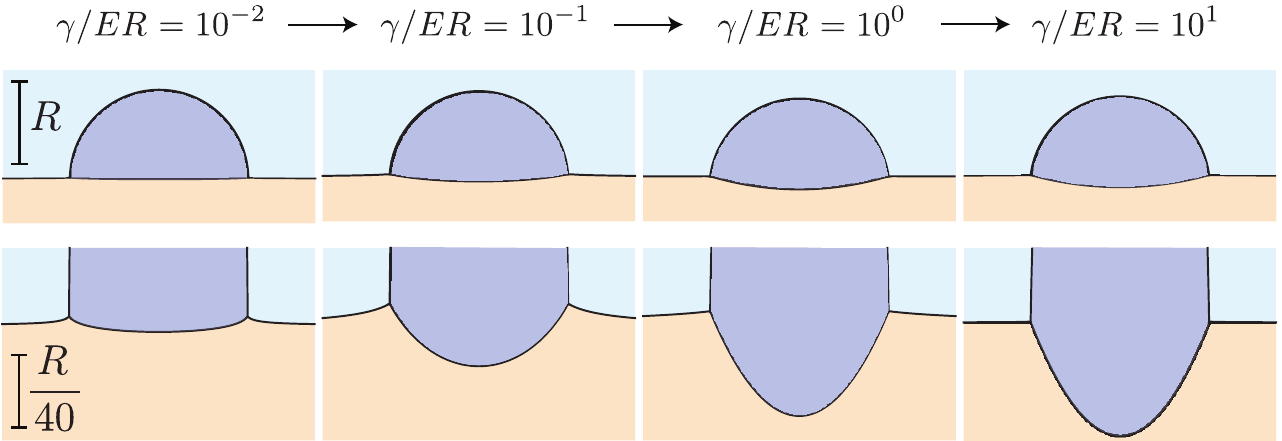}
\caption{
Simulated shapes of the drop (dark blue) and of the solid substrate (light orange) for different values of $\gamma/ER$, $\gamma/\gamma_s=0.3$, $\tilde a=0$. Top:~As the solid becomes softer (increasing $\gamma/ER$) the drop sinks increasingly deeper in the substrate. Bottom: Zoom near the contact line region, with the vertical scale stretched 40 times. The elevation at the contact line first increases for increasing softness, but on further increasing the softness the elevation subsequently decreases.
\label{fig:shapes}}
\end{figure}
In the remainder we rescale all lengths by $R$ and the pressure by $\gamma/R$. The macroscopic model depends on two dimensionless parameters: $\gamma/(ER)$ and $\gamma/\gamma_s$. The microscopic model has an additional parameter, $\tilde{a}=a/R$, or equivalently $\gamma/(Ea)$. 
Solving for the profile in Fourier space, $\hat h(q)$,  [Eq.~(\ref{eq:sigmadrop})] gives:
\begin{equation}\label{eq:hq}
\hat{h}(q) =\frac{\gamma}{ER} \sin \theta 
\left\{ \frac{ \hat{f}_n(q,\tilde{a})\, \hat{\cal K}(q)}{1+  \frac{\gamma}{ER} \frac{\gamma_s}{\gamma} q^2 \hat{\cal K}(q)}
\right\},
\end{equation}
where
\begin{equation}
\hat{f}(q,\tilde{a}) = 2 \left(e^{-\frac{1}{2}(\tilde{a}q)^2}\cos q - \frac{\sin q}{q}\right) , \quad \mathrm{and} \quad \widehat{\cal K}(q)\equiv \widehat{Q}^{-1}(q) = \frac{3}{2|q|}.
\end{equation}
The macroscopic theory corresponds to $\tilde{a}=0$. Numerical solutions for $h(x)$ that satisfy (\ref{eq:horneumann}) are obtained iteratively, by adjusting $\theta$. Typical results are shown in Fig.~\ref{fig:shapes}. The assumption $\gamma_{SV}=\gamma_{SL}$ makes the contact line left-right symmetric in the rigid limit. For  $\gamma_{SV}\neq\gamma_{SL}$ this symmetry is broken, but results will be qualitatively the same.

\section{Numerical results}

\subsection{Microscopic and macroscopic contact angles: two transitions}

\begin{figure}
\includegraphics{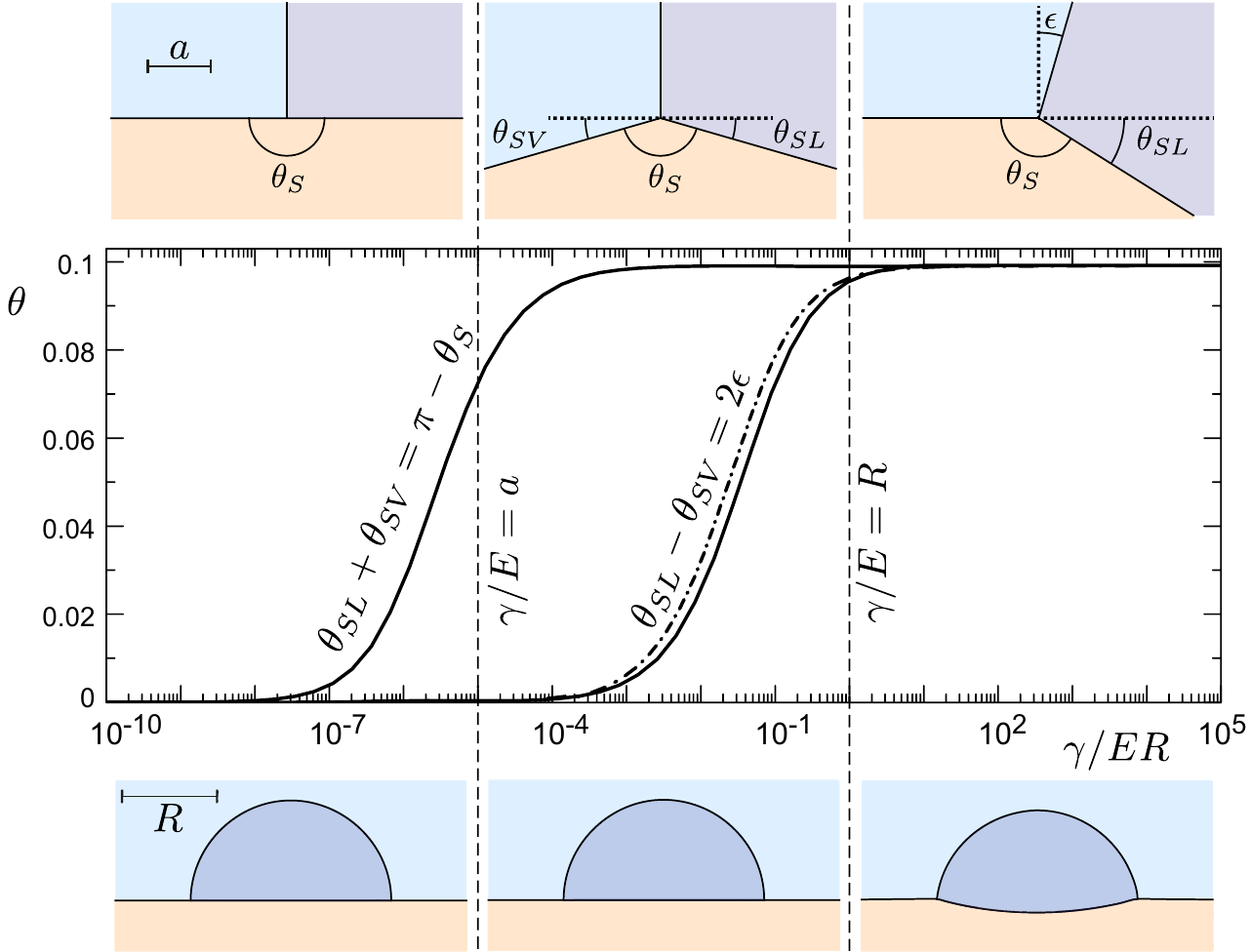}
\caption{Double transition of contact line angles, for 2D drops (solid) and 3D drops (dashed) The stiffness is varied by  the softness parameter $\gamma/ER$, at fixed $\gamma/\gamma_s = 0.1$ and $\tilde{a}=10^{-5}$. 
The left curve give the deviation of the solid angle, $\pi-\theta_S$, showing the onset of the cusp.
The right hand curves give the cusp rotation angle $\epsilon$, defined from the microscopic wetting angles (top panel, scale $a$).  The bottom panel shows the corresponding macroscopic drops shapes. This shows that (i) the microsocopic geometry of the contact line first develops a Neumann-like cusp when $\gamma/E$ is of the order of the microscopic scale $a$ and (ii) the macroscopic angle of the drop is altered only when $\gamma/E$ reaches the size of the drop $R$. 
\label{fig:doubletransition}}
\end{figure}

We now describe how the wetting angles depend on the stiffness of the substrate. It is important to distinguish between the microscopic wetting angles $\theta_S$, $\theta_{SL}$, $\theta_{SV}$ (Fig.~\ref{fig:doubletransition}), and the macroscopic angle $\theta$ (Fig.~\ref{fig:sketch}a). The former angles reveal the the microscopic geometry of the contact line region, while the latter is the apparent angle of the spherical cap with respect to the undeformed substrate. The results presented in Fig.~\ref{fig:doubletransition} are obtained for $\gamma/\gamma_s=0.1$, $\tilde{a}=10^{-5}$, and Young's angle $\theta_Y=90^\circ$ (since $\gamma_{SV}=\gamma_{SL}$).

Our key finding is that the wetting angles undergo two transitions when changing the softness parameter: (i) the \emph{microscopic} angles evolves continuously from Young's law (rigid limit) to Neumann's law (soft limit), controlled by the parameter $\gamma/Ea$, (ii) the \emph{macroscopic} apparent contact angle $\theta$ evolves continuously from Young's law (rigid limit) to Neumann's law (soft limit), controlled by the parameter $\gamma/ER$. As discussed in more detail below, these soft and rigid limits are known analytically, while for the transition regions described in Fig.~\ref{fig:doubletransition} we rely on numerical solution of (\ref{eq:hq},\ref{eq:horneumann}). 

For $\gamma/Ea \ll 1$, in the perfectly rigid limit, the microscopic solid angle is unchanged, {\em i.e.}, $\theta_S=180^\circ$, while the liquid angle $\theta=\theta_Y$ is equal to Young's angle. At the first transition, $\gamma/Ea \sim 1$, the solid angle changes and the geometry evolves towards a Neumann cusp. This transition is quantified in the plot in Fig.~\ref{fig:doubletransition}, where the solid line on the left measures the solid deflection $\pi-\theta_S = \theta_{SV}+\theta_{SL}$, for different values of softness. In this regime the macroscopic angle $\theta=\theta_Y=90^\circ$ does not change, and hence the geometry is fully characterized by the single angle $\theta_S$. The Neumann cusp, as predicted in the macroscopic theory for all values of the stiffness, is only recovered for $\gamma/Ea \gg 1$.

The second transition does involve the apparent contact angle $\theta$. As long as $\gamma/ER \ll 1$, this angle is unaffected by elastic deformations, as these are localized only in a narrow zone near the contact line. The macroscopic angle changes only once the scale of the deformation $\gamma/E$ becomes comparable to the drop size $R$, {\em i.e.} when $\gamma/ER \sim 1$. In the very soft limit the angle saturates, reaching the value expected for a liquid lens. As is visible in the sketches in Fig.~\ref{fig:doubletransition}, the solid-vapor angle $\theta_{SV}=0$ in this limit, consistent with previous predictions \citep{Style12,Limat12}. Microscopically, the geometry of the contact angles is completely specified by the amount of rotation of the Neumann angles, as is indicated by the angle $\epsilon$ (Fig.~\ref{fig:doubletransition}, top). The transition is quantified by the solid line in the plot of Fig.~\ref{fig:doubletransition} (middle panel), where we observe how the rotation angle $\epsilon$ evolves from the limits corresponding to Young and Neumann respectively. Indeed, the crossover is observed around $\gamma/ER \sim 1$.

For completeness, we also carried out the macroscopic analysis for three-dimensional axisymmetric drops. The analysis is analogous to that in the two-dimensional case, except that we use the Hankel transform \citep{Style12}. As we here self-consistently determine the value of the angle $\theta$, we are for the first time able to assess the validity of the ``elastic line tension" argument~\citep{White03,Style12}. The result for the deflection angle $\epsilon$ is shown as the dashed line in Fig.~\ref{fig:doubletransition}, from which it is apparent that this angle is almost indistinguishable from that obtained in the two-dimensional case. 
The trend for $\theta$, decreasing when reducing the stiffness, is {\em opposite} to that predicted by considering a positive line tension, but in agreement with recent experiments (Fig. 3a of \cite{Style13}).
\begin{figure}
\includegraphics{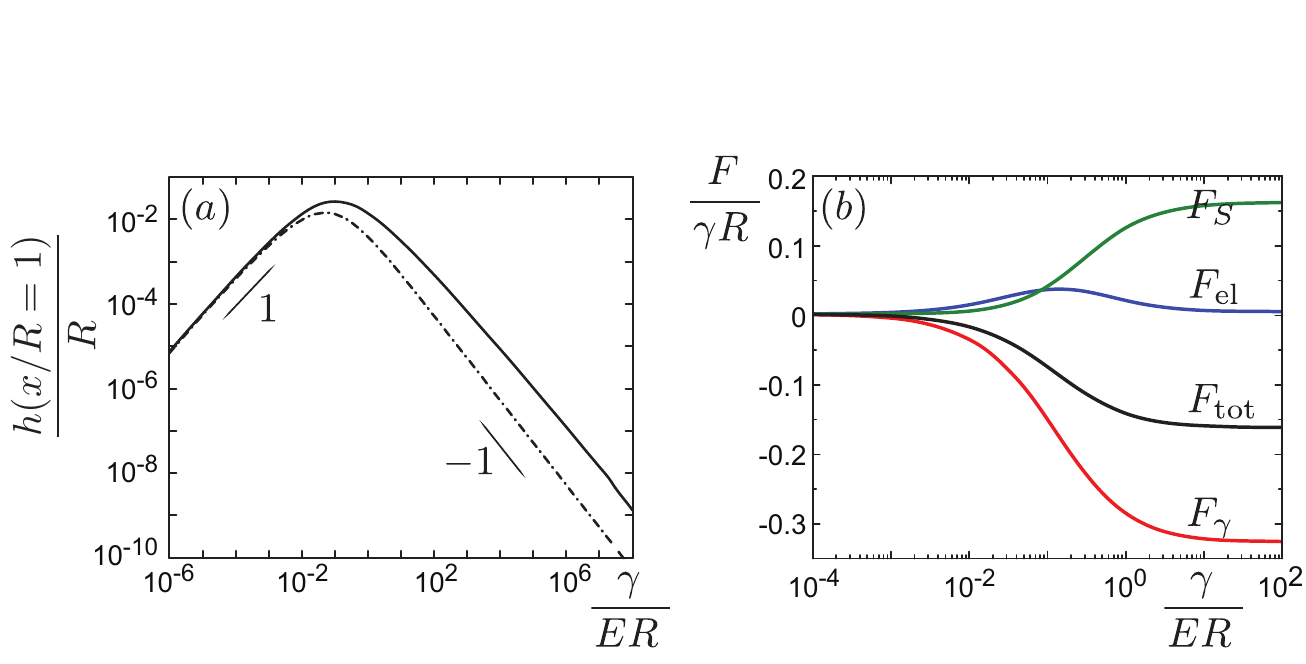} 
\caption{(a) Elevation at the contact line as function of $\gamma/ER$ for $\gamma/\gamma_s=0.1$: solid
lines are results from 2-D theory, dash-dotted lines from 3-D theory. Deformations at the contact line are nonmonotonic, as seen also in Fig.~\ref{fig:shapes}. (b) Elastic energy $\mathcal{F}_{el}$, capillary solid energy $\mathcal{F}_S$, and capillary liquid energy $\mathcal{F}_\gamma$, as function of $\gamma/ER$ for $\gamma/\gamma_S = 0.5$. The total free energy decreases as the softness increases. Reference energy ($E=0$) is a drop on rigid solid. 
\label{fig:ee}}
\end{figure}

\subsection{Elevation of the contact line and total energy vs. softness}
While varying $\gamma/ER$ we observe that the elevation of the contact line for the macroscopic model ($\tilde{a}=0$) evolves non-monotonically (cf. Fig.~\ref{fig:shapes}). This is quantified in Fig.~\ref{fig:ee}, showing the contact line elevation $h(x/R=1)$ as a function of $\gamma/ER$: The contact line elevation $h(1)$ has a maximum for $\gamma/ER\sim 0.1$. In the stiff limit ($E\to\infty$) the solid opposes any stress without deforming while in the soft limit ($E\to 0$) the geometry of the drop-substrate interface is identical to that of a floating liquid lens. From the numerical results we find that the contact line height fits closely to
\begin{eqnarray}\label{eq:as}
h(x/R=1) \sim  \left(\frac{\gamma}{ER}\right) \ln \left(\gamma/ER\right), \quad \mathrm{and} 
\quad h(x/R=1) \sim \left( \frac{\gamma}{ER}\right)^{-1} \ln \left(ER/\gamma\right),
\end{eqnarray}
for small and large $\gamma/ER$, respectively. The rigid limit was previously found by \citet{Limat12}, who showed that for weak deformations the solid surface tension provides a natural regularization of the logarithmic divergence at the contact line. The dashed line (three-dimensional drop), suggests a $h\sim (\gamma/ER)^{-1}$ dependence for $\gamma/ER\gg1$. 

Another characteristic of the deformation is the central deflection, $h(x/R=0)$. As seen in Fig.~\ref{fig:shapes}, this deflection increases monotonically with softness, starting from $0$ (drop on rigid solid) and saturating at the value for a floating lens. The crossover between these limits is again governed by the macroscopic length, and lies around $\gamma/ER \sim 0.1$.

Finally, we calculate the elastic energy ($\mathcal{F}_{el}$), the solid capillary energy ($\mathcal{F}_S$), and the liquid capillary energy ($\mathcal{F}_\gamma$) for a droplet on a substrate of varying stiffness [see Fig.~\ref{fig:ee}(b)]. In these calculations, the droplet volume was kept constant. This allows for a direct comparison with recent experiments by \citet{StylePNAS13} where droplets were found to move to softer regions on a substrate with a stiffness gradient. Our calculations, which predict a lower total energy $\mathcal{F}_{tot}\equiv \mathcal{F}_{el}+\mathcal{F}_S+\mathcal{F}_\gamma$ for droplets on softer substrates, provide an explanation for this  observation. The energies $\mathcal{F}_{el}$ and $\mathcal{F}_S$ are actually increasing for softer surfaces, but by a smaller amount than the gain in liquid-vapor energy ($\mathcal{F}_\gamma$).

\section{Discussion}

In this work, we have calculated the shapes of drops on soft substrates by minimization of the elastic and capillary energies. We find that in addition to the drop size $R$ and the elastocapillary length $\gamma/E$, a third length-scale is required in order to fully describe elastocapillary interactions: The molecular scale $a$. This extra length scale is crucial for describing the Young to non-Young transition that occurs for sessile droplets on progressively softer substrates. In addition, the regularization is necessary to avoid a singularity of elastic stress at the contact line. Such a singular tendency implies that, generically, the substrates will be deformed beyond the linear elastic regime --~one expects strain hardening and even plasticity \citep{Limat12}~-- which up to present have never been taken into account. In addition, the elastic free energy posed in (\ref{eq:Fel}) is strictly speaking only valid for small slopes of the solid substrate; this is why our numerical results in Fig.~\ref{fig:shapes} were chosen at moderate $\gamma/\gamma_s$. For larger slopes one in principle has to deal with this ``geometric" nonlinearity, even when assuming that  the rheology of the elastic solid remains linear. It is not clear a priori whether this would lead to an elastic contribution to the Neumann boundary condition for soft solids.

In order to derive the present results, we have left out the second difficulty of the problem, namely  the strain-dependence of surface free energies (Shuttleworth effect). Such a strain-dependence has been shown to lead to significant tangential elastic stress \citep{WAS13}. Generically, this induces tangential displacements that are comparable in magnitude to the normal deflection of the substrate, hence changing the energetics of the problem. We emphasize again that this effect is of primary importance for \emph{quantitative} comparison with experiments; the more so since tangential stresses are expected to give an elastic contribution to the contact angle boundary condition \citep{MDSA12b}. Qualitatively, however, the present model does capture two key experimental observations: softer substrates lead to a decrease of macroscopic contact angle, and to a decrease in total free energy. The derivation of the contact angle selection and substrate deformation including all difficulties mentioned above (nonlinearities and Shuttleworth effect) is the following major challenge for the field.

%

\end{document}